\documentclass[twocolumn,preprintnumbers,amsmath,amssymb,prl]{revtex4-1}
\usepackage{graphicx}% Include f files
\usepackage{dcolumn}% Align table columns on decimal point
\usepackage{bm}% bold math
\usepackage{units}%
\usepackage{stmaryrd}%
\usepackage[breaklinks=true,pdfborder={0 0 0}]{hyperref}
\usepackage[usenames]{color}
\usepackage{booktabs}
\usepackage{chemformula}
\usepackage{tikz}
\usepackage{multirow}
\usepackage{mathtools,etoolbox}
\DeclarePairedDelimiterX{\abs}[1]{\lvert}{\rvert}{\ifblank{#1}{{}\cdot{}}{#1}}
\usepackage{physics}
\hypersetup{colorlinks=true,citecolor=red}
\usepackage[usenames]{color}
\newcommand{\rrr}[1]{{\color{red}#1}}

\begin{document}
\newlength{\LL} \LL 1\linewidth
\title{Optimized Quantum Drude Oscillators for Atomic and Molecular Response Properties}
\author{Szabolcs G\'oger}
\author{Almaz Khabibrakhmanov}
\author{Ornella~Vaccarelli}
\author{Dmitry~V.~Fedorov}
\author{Alexandre~Tkatchenko}
\email[E-mail: ]{alexandre.tkatchenko@uni.lu}
\affiliation{Department of Physics and Materials Science, University of Luxembourg, L-1511 Luxembourg City, Luxembourg}
%\date{\today}

\begin{abstract}
The quantum Drude oscillator (QDO) is an efficient yet accurate coarse-grained approach that has been widely used to model electronic and optical response properties of atoms and molecules, as well as polarization and dispersion interactions between them. Three effective parameters (frequency, mass, charge) fully characterize the QDO Hamiltonian and are adjusted to reproduce response properties. However, the soaring success of \emph{coupled} QDOs for many-atom systems remains fundamentally unexplained and the optimal mapping between atoms/molecules and oscillators has not been established. Here, we present an optimized parametrization (OQDO) where the parameters are fixed by using only dipolar properties.
For the periodic table of elements as well as small molecules, our OQDO model accurately reproduces atomic (spatial) polarization potentials and multipolar dispersion coefficients, elucidating the high promise of the presented model in the development of next-generation quantum-mechanical force fields for (bio)molecular simulations.
\end{abstract}

\maketitle

The development of predictive model Hamiltonians that can describe various properties of realistic molecules and materials is a cornerstone of modern physics~\cite{Martyna-RMP} and chemistry~\cite{Prezhdo-ChemRev}. The quantum Drude oscillator (QDO) is arguably the most powerful Hamiltonian for accurate and efficient modeling of atomic and molecular response~\cite{Hermann,Martyna-RMP,Wang2001,Sommerfeld2005,Jones2013,Sadhukhan2016,Lemkul2016, Schroeder2010, Anisimov2005}. Within the coarse-grained QDO model, the response of valence electrons is described \emph{via} a quasi-particle \emph{drudon} with a negative charge $-q$ and mass $\mu$, harmonically bound to a positively-charged pseudo-nucleus of charge $q$ with a characteristic frequency $\omega$. The many-body extension of the QDO model (the \emph{coupled} QDO model) has been widely employed to study both molecules and materials, including their electronic~\cite{THOLE1981,Hermann2017} and optical~\cite{Ambrosetti2022} properties, polarization~\cite{Whitfield2006,Jordan-pol}, dispersion~\cite{Bade1957, Bade2, Whitfield2006, Tkatchenko2012, Reilly2015, Sadhukhan2016, Sadhukhan2017,Ambrosetti2014, DiStasio2014, Stoehr2021,Karimpour2022}, and exchange~\cite{Vaccarelli2021, Fedorov2018,Tkatchenko2021} interactions, as well as a wealth of non-additive field effects in quantum mechanics~\cite{Kleshchonok2018, Stoehr2021} and quantum electrodynamics~\cite{Karimpour2022_JPCL, Karimpour2022}. Coupled QDOs are also extensively used in the development of van der Waals (vdW) density functionals~\cite{Tkatchenko2009, Tkatchenko2012,Ambrosetti-QDO}, quantum mechanical~\cite{Jones2013, Martyna-RMP} and polarizable force fields~\cite{Harder2006,Lopes2013, Adluri2015, Sokhan2015, Piquemal-MBD1} as well as recent machine learning force fields~\cite{Heikki2021,Piquemal-MBD2}. Despite such a wide applicability of the coupled QDO model, its success in describing real atoms remains fundamentally unexplained and the optimal mapping between atoms and oscillators has not been established. In this Letter, we develop an optimized parametrization (OQDO) where the parameters are fixed by using only the well-known atomic dipolar properties. Remarkably, OQDO reproduces spatial atomic polarization potentials and atomic multipolar dispersion 
coefficients. Our OQDO model for atoms and small molecules also paves the way to develop next-generation quantum-mechanical force fields for (bio)molecular simulations.

\begin{figure*}[t]
\includegraphics[width=0.8\linewidth]{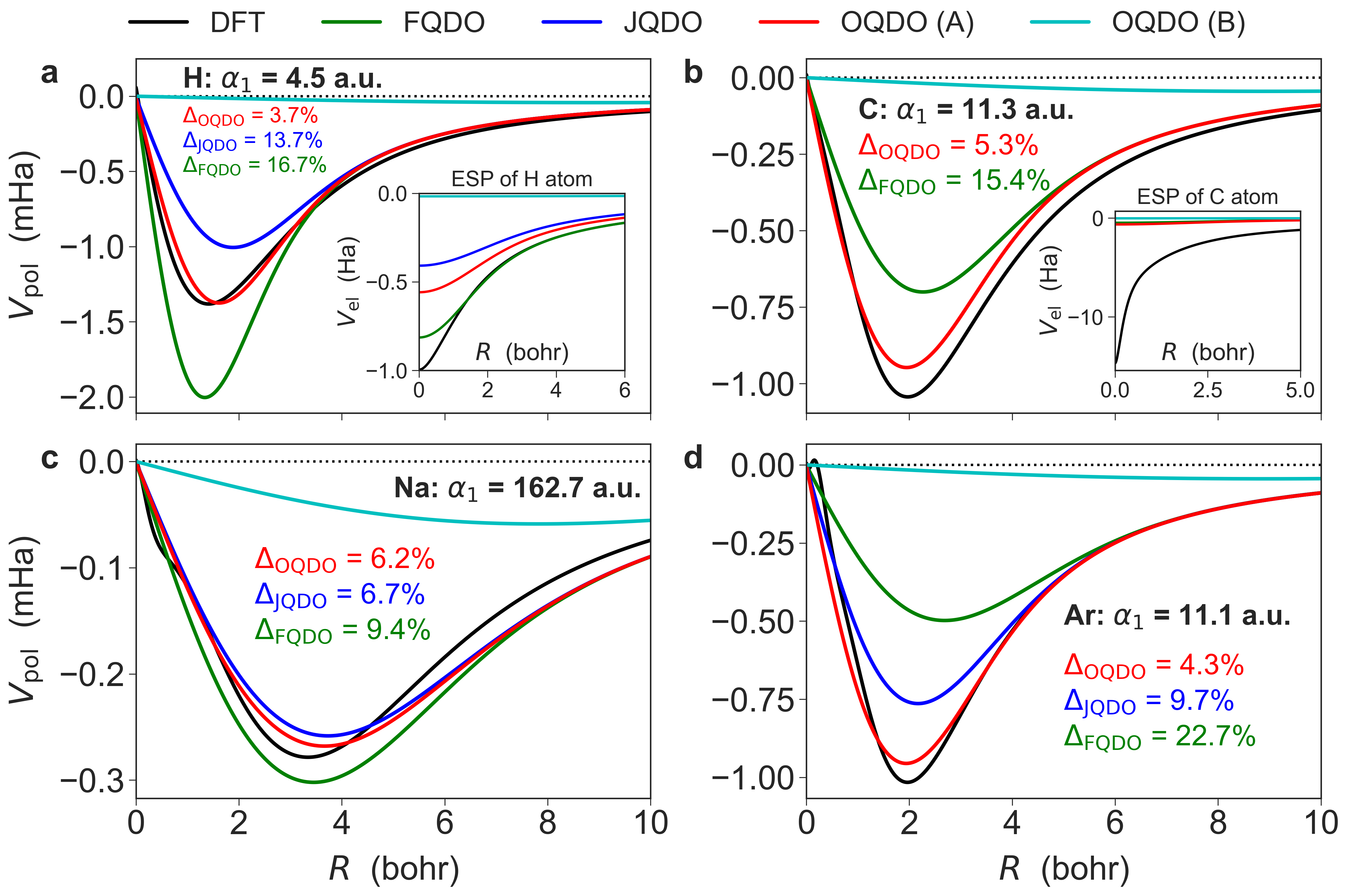}
\caption{Polarization potential curves $V_{\mathrm{pol}}(\mathbf{r})$ calculated with DFT-PBE0 and various QDO parametrizations for (a) hydrogen, (b) carbon (no JQDO values are available), (c) sodium, and (d) argon atoms. 
The FQDO and JQDO parametrization schemes are described by Eqs.~\eqref{eq:fqdo} and \eqref{eq:jqdo}, respectively.
OQDO(A) and OQDO(B) correspond to the two solutions of the transcendental equation given by Eq.~\eqref{eq:newQDOparam}.
In all cases the direction along the applied field was chosen for the plots. The reference values for dipole polarizability $\alpha_1$ are shown for each element. The numerical values of the normalized root-mean-squared error ($\Delta$) are displayed for the three QDO flavors. For hydrogen and carbon atoms, the unperturbed electrostatic potentials (ESP) $V_{\mathrm{el}}(\mathbf{r})$ are shown as insets, indicating that a QDO captures the response of atomic electron density, but not the static potential itself.}
\label{fig:QDO_response}
\end{figure*}

The three parameters \{$q, \mu, \omega$\} fully define the QDO, and three atomic response properties could be chosen to fix them, meaning that the choice of QDO parameters is not unique. In addition, all QDO response properties -- multipolar polarizabilities and dispersion coefficients -- are uniquely fixed by the three parameters \emph{via} closed-form relations~\cite{Jones2013}. The static dipole polarizability of a QDO, $\alpha_1 = q^2/\mu\omega^2$, conveniently combines all three parameters, and it is natural to set this expression to the reference atomic $\alpha_1$. The QDO expression for the dipole-dipole dispersion coefficient $C_6 = \frac{3}{4} \hbar\omega\alpha_1^2$ is identical to the London formula and allows to fix $\omega$ if the reference atomic values of $C_6$ and $\alpha_1$ are given. Since $\alpha_1$ and $C_6$ are accurately known for all elements in the periodic table~\cite{Gobre2016,Gould2016,Database2018}, they form a baseline for the QDO parametrization. However, one more condition is required to obtain \{$q, \mu, \omega$\}, for which different constraints can be imposed. A reasonable idea is to fix $q = 1$ a.u., since a QDO should reproduce the response of electrons. This results in the fixed-charge QDO 
(FQDO)
\begin{equation}
\label{eq:fqdo}
q = 1 \; ,~~\omega = {4C_6}/{3\hbar\alpha_1^2} \; ,~~\mu = {9\hbar^2\alpha_1^3}/{16C_6^2} \quad .
\end{equation}
However, fixing $q$ and using QDO recursion relations for high-order response usually yields large errors in the multipolar response properties (see Fig.~2 and Refs.~\cite{XDM,D4}). A more rigorous approach was suggested by Jones \emph{et al.}~\cite{Jones2013} by employing the dipole-quadrupole dispersion coefficient $C_8$. The mapping $\{\alpha_1, C_6, C_8\} \rightarrow \{q, \mu, \omega\}$ yields the Jones QDO (JQDO) parametrization scheme
\begin{equation}
\label{eq:jqdo}
q = \sqrt{\mu\omega^2\alpha_1} \; ,~~\omega = {4C_6}/{3\hbar\alpha_1^2} \; ,~~\mu = {5\hbar C_6}/{\omega C_8}\ .
\end{equation}
The JQDO approach improves the multipolar response
over the FQDO model, while simulations using the coupled JQDO model captured many remarkable properties of bulk water and its surface~\cite{Martyna-water,Sokhan2015}. However, the $C_8$ dispersion coefficient is not directly measurable, and accurate \emph{ab initio} calculations of quadrupole ($\alpha_2$) and octupole ($\alpha_3$) polarizabilities and $C_8$ -- $C_{10}$ dispersion coefficients are currently technically feasible only for closed-shell species (noble-gas atoms and small molecules) or alkali and alkaline-earth atoms with $s$ valence shells~\cite{Porsev2003,Porsev2006,JIANG2015,Tao2016}. For other open-shell atoms (containing $p$, $d$, or $f$ valence shells), convergence of \emph{ab initio} response calculations becomes a technical hurdle~\cite{Rob-privatecomm}. 
Thus, using higher-order atomic response properties does not lead to a parametrization that would be universally applicable across the periodic table as well as for small molecules.

Here, we introduce an optimized QDO parametrization (OQDO), where we effectively map dipolar atomic quantities \{$\alpha_1, C_6$\} to the oscillator parameters. The third parameter is fixed by using the force balance equation for vdW-bonded dimers derived recently~\cite{Fedorov2018,Vaccarelli2021,Tkatchenko2021}. Two equations for $q$ and $\omega$ follow the JQDO scheme, whereas the third one is replaced with a transcendental equation for a product $\mu\omega$ to be solved numerically (\emph{vide infra})
\begin{equation}
\label{eq:newQDOparam}
\mu = \frac{5 \, \hbar \, {\rm C}_6 }{ \omega \, {\rm C}_8 } \rightarrow  \exp( \frac{2\mu\omega R_{\rm vdW}^2}{\hbar} ) = \frac{2^7 \cdot (\alpha_{\rm fsc}^{-1/3} a_0)^4}{(3 \hbar / \mu\omega )^2} \; ,
\end{equation}
where $\alpha_{\rm fsc} = e^2 / 4\pi\varepsilon_0 \hbar c$ is the fine-structure constant,
and the vdW radius ($R_{\rm vdW}$) is calculated \emph{via} the universal formula 
connecting it with the dipole polarizability
\begin{equation}
\label{eq:dip_pol_RvdW_cf}
\alpha_1 (R_{\rm vdW}) = {(4\pi\varepsilon_0)\, R_{\rm vdW}^7}/{(\alpha_{\rm fsc}^{-1/3} a_0)^4} \ ,
\end{equation}
as obtained from the balance between exchange repulsion
and vdW dispersion attraction forces at the equilibrium distance in homonuclear atomic dimers~\cite{Fedorov2018,Vaccarelli2021}.

\begin{figure*}[t]
\includegraphics[width=0.85\linewidth]{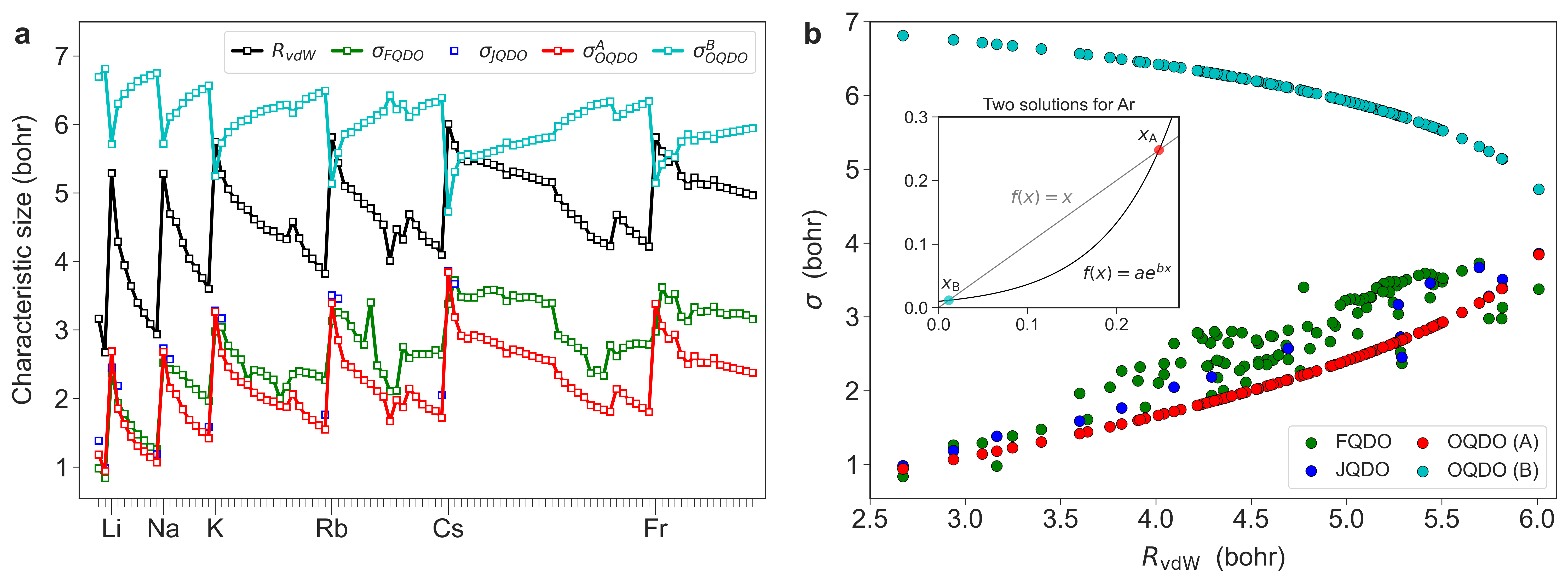}
\caption{(a) Periodic variations of the QDO length $\sigma = \sqrt{\hbar/2\mu\omega}$ with the atomic number for the three different parametrizations, as compared to the atomic vdW radii ($R_{\mathrm{vdW}}$) which are evaluated \emph{via} Eq.~\eqref{eq:dip_pol_RvdW_cf} using the reference atomic polarizabilities~\cite{Gobre2016}. (b) Correlation between $R_{\mathrm{vdW}}$ and $\sigma$ (within the three QDO parametrizations) for 102 elements in the periodic table. The schemes OQDO(A) and OQDO(B) correspond to two solutions of Eq.~\eqref{eq:TranscEq}, as illustrated using the example of Ar in the inset.}
\label{fig:Sigma_Rvdw}
\end{figure*}

The predictive power of the QDO Hamiltonian arises from its ability to effectively model integrated atomic response. However, when modeling a molecule or a solid, coupled QDOs must capture interactions between atoms. Considering two atoms $i$ and $j$ and using interatomic perturbation theory~\cite{stone2013theory,Szalewicz-ChemRev}, the interaction energy can be written as the integrated product of the atomic electron density of $i$ with the electric potential of atom $j$~\cite{kaplan2006intermolecular, stone2013theory}    
\begin{equation}
\label{eq:E_int}
E_{\mathrm{int}}=\int \rho_{\rm i}(\mathbf{r}) V_{\rm j}( \mathbf{r}) \dd^3  \mathbf{r} \ .
\end{equation}
This formula is valid for all dominant noncovalent interactions: electrostatics, induction/polarization, exchange-repulsion, and dispersion. The validity is evident for the former two cases~\cite{kaplan2006intermolecular,stone2013theory}, and it was shown that dispersion~\cite{Lilienfeld2004} and exchange~\cite{VanVleet2016} interactions can also be \emph{effectively} represented using the form of Eq.~(\ref{eq:E_int}). 

Response properties are given \rrr{by} variations of $E_{\rm int}$ \rrr{as}
\begin{equation}
\label{eq:dE_int}
\delta E_{\mathrm{int}} = \int \left(  \delta \rho_{\rm i}(\mathbf{r}) V_{\rm j}(\mathbf{r}) + \rho_{\rm i}( \mathbf{r}) \delta V_{\rm j}(\mathbf{r}) \right) \dd^3  \mathbf{r} \; .
\end{equation}
For an external electric field $\mathbf{E}$, which can also model the effect of environment, $\delta \rho (\mathbf{r}) = \rho_{E} (\mathbf{r}) - \rho (\mathbf{r})$, where $\rho_{E}(\mathbf{r})$ is the electron density under the external field. Then, the dominant contribution to $\delta V_{\rm j}( \mathbf{r})$ is given by the polarization potential~\cite{Comment}
\begin{equation}
\label{eq:vpol_def}
V_{\mathrm{pol}}(\mathbf{r}) = -\frac{1}{4\pi \varepsilon_0} \int  \frac{\rho_{E}(\mathbf{r'})-\rho(\mathbf{r'})}{ |\mathbf{r-r'}|} \dd^3  \mathbf{r'}  \;,
\end{equation}
which describes the change in
the electrostatic potential of the system due to the polarization of its charge density by the presence of another moiety (an electric field in this case). For the QDO, the integral in Eq.~\eqref{eq:vpol_def} can be evaluated analytically, yielding the following expression
\begin{equation}
\label{eq:vpol_qdo}
V_{\mathrm{pol}}^{_{\mathrm{QDO}}}(\mathbf{r}) = \frac{-q}{4\pi \varepsilon_0} \left( \frac{\text{erf}(\tilde{r}/\sigma \sqrt{2})}{\tilde{r}} - \frac{\text{erf}(r/\sigma \sqrt{2})}{r} \right) \;,
\end{equation}
where $\tilde{\mathbf{r}} = \mathbf{r} + \alpha_1 \mathbf{E}/q$ is the field-induced oscillator coordinate and $\sigma = \sqrt{\hbar/2\mu\omega}$ is the QDO spread~\cite{Szabo2022}.
In the Supplemental Material (SM)~\cite{SM}, we present 
$V_{\mathrm{pol}}^{_{\mathrm{QDO}}}(\mathbf{r})$ in comparison to
$V_{\mathrm{pol}}(\mathbf{r})$ calculated for 21 atoms (between H -- Ca and Kr) within hybrid density-functional theory DFT-PBE0~\cite{Carlo1999,PBE,qchem,Multiwfn,ESP} shown to yield a highly accurate description of electronic response~\cite{Hait2018} comparable to coupled-cluster calculations~\cite{SM}.
Here, we remark that the strength of the electric
field was chosen individually for each element depending on its reference static dipole 
polarizability~\cite{Gobre2016,Database2018} so that the  
field-induced dipole moment is set as $\mathbf{d} = \alpha_1 \mathbf{E}= 0.01$~a.u., for all atoms~\cite{CCC}.

Before comparing $V_{\mathrm{pol}}({\bf r})$ for real atoms with different QDO flavors, it is instructive to consider which atomic properties can be faithfully captured by a QDO. Firstly, the QDO does not aim to describe static properties of the atomic electron density, but rather its response under applied static and fluctuating fields, as demonstrated also by the insets in Fig.~\ref{fig:QDO_response}a,b. The electrostatic potential (ESP) of a QDO is given by $V_{\rm el}^{_{\rm QDO}} = -q \cdot {\rm erf}(r/\sigma \sqrt{2})/r$, so the charge $q$ determines its magnitude. This explains why $V_{\rm el}^{_{\rm FQDO}}$ yields a good agreement with $V_{\rm el}^{\rm DFT}$ for hydrogen. However, the QDO model does not describe $V_{\rm el}$ for many-electron atoms because $q \sim 1$ a.u., while the ESP of atoms scales non-linearly with $Z$ (see the example of carbon in the inset of Fig.~\ref{fig:QDO_response}b). Secondly, the harmonic response captured by a QDO model should be sufficient to accurately describe integrated electronic displacements induced by weak fields. However, it is much less clear how well different QDO parameterizations perform for distributed polarization potentials described by Eq.~\eqref{eq:vpol_def} for many-electron systems, given the analytical form of $V_{\mathrm{pol}}^{_{\mathrm{QDO}}}(\mathbf{r})$ in Eq.~\eqref{eq:vpol_qdo}. To answer this question,
in Fig.~\ref{fig:QDO_response} we compare $V_{\mathrm{pol}}$ of real atoms and $V_{\mathrm{pol}}^{_{\mathrm{QDO}}}(\mathbf{r})$ employing the three QDO models discussed above.
We used the accurate \emph{ab initio} reference data on $\alpha_1$ and $C_6$~\cite{Gobre2016,Database2018,Derevianko2010} to parametrize FQDO and OQDO. When available, we also used the analogous data on $C_8$~\cite{Porsev2003,Porsev2006,JIANG2015} to parametrize the JQDO model. We observe that the OQDO model is able to reproduce the full range of the polarization potential of real atoms with a reasonable accuracy, showing a significantly better agreement with the DFT-PBE0 results than FQDO and JQDO. To quantify this, for each atom we calculated the root-mean-squared-error (RMSE) of the three QDO curves with respect to the PBE0 reference curves and normalized the RMSE using the equilibrium depth of the PBE0 curve. The OQDO flavor has an error of 8.9\% when averaged over 21 atoms, whereas JQDO and FQDO yield average errors of 13.2\% and 15.4\%, respectively. We also emphasize that the predictions of the OQDO model remain accurate for many-electron atoms such as noble gases and alkali metals. 
It is especially reassuring that the OQDO model reproduces the nonlinear $V_{\rm pol}(\mathbf{r})$ curves obtained from DFT calculations without any adjustments. In fact, the OQDO performance is sensitive to variations in the QDO parameters 
(solutions A or B in Fig.~\ref{fig:QDO_response}),
so the satisfactory agreement shows that the chosen OQDO(A) model accurately describes real atoms. 
The significant differences between the predictions of various parametrizations for $V_{\mathrm{pol}} (\mathbf{r})$ underline the importance of the optimal mapping between atomic response properties and QDO parameters.

\begin{figure}[t]
\includegraphics[width=0.9\linewidth]{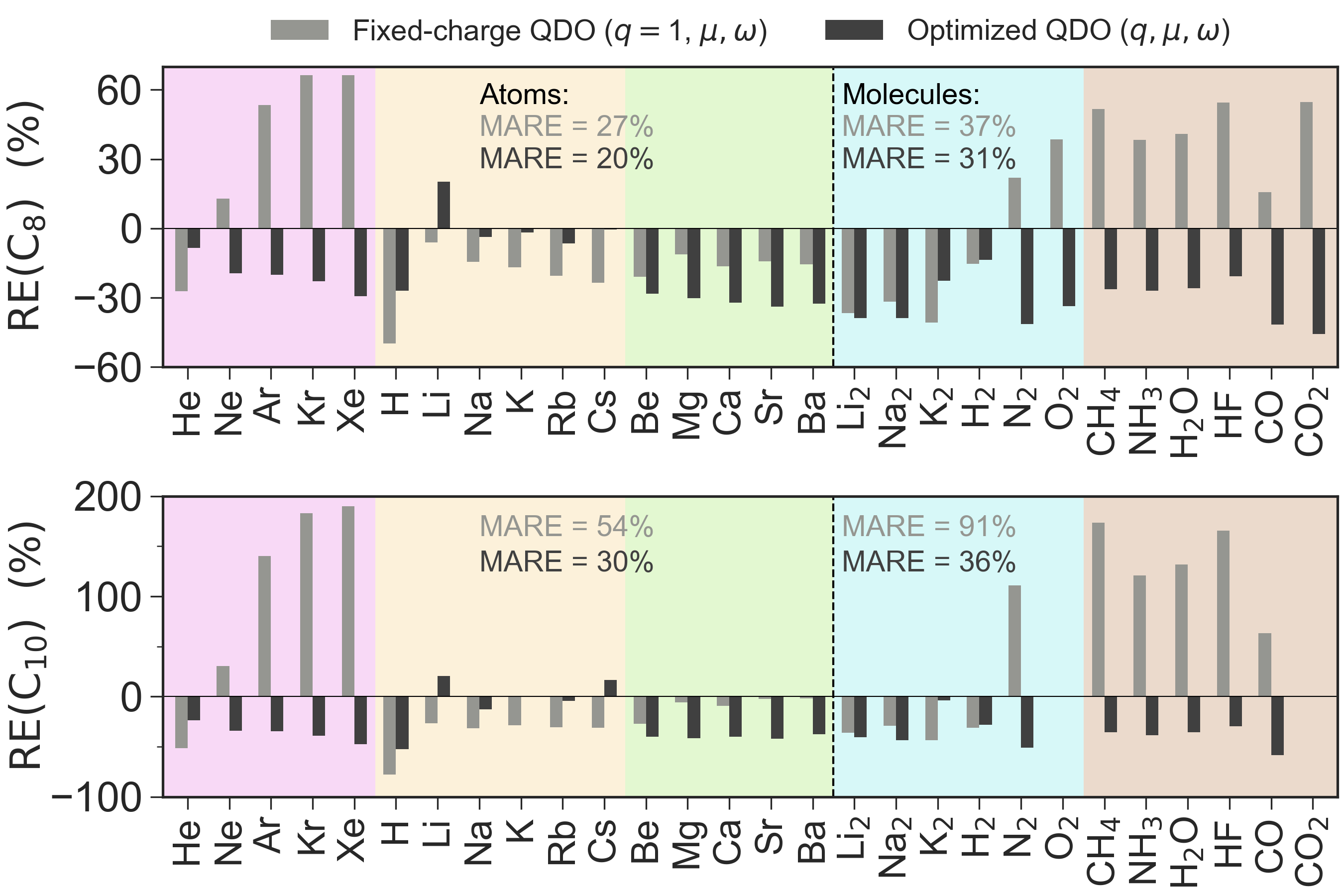}
\caption{Multipolar dispersion coefficients $C_8$ and $C_{10}$ as predicted by FQDO ($q=1$) and OQDO models. Relative error ${\rm{RE}} = (C_j-C_j^{\rm{ref}})/C_j^{\rm{ref}}$ with respect to \emph{ab initio} reference data~\cite{Porsev2003, Porsev2006, JIANG2015, Tao2016} is plotted. For the two models, numerical values of mean absolute relative errors (MARE) are evaluated separately for atoms and molecules. In the case of $\mathrm{O_2}$ and $\mathrm{CO_2}$, no reliable \emph{ab initio} reference data for $C_{10}$ could be found.}
\label{fig:QDO_disp}
\end{figure}

We discuss now the technical aspects of deriving the two solutions of the OQDO model (see OQDO(A) and OQDO(B) in Fig.~\ref{fig:QDO_response}), and their connection to real atoms.  The starting point is Eq.~\eqref{eq:dip_pol_RvdW_cf} that connects the atomic vdW radius and its dipole polarizability. Within the QDO model, Eq.~\eqref{eq:dip_pol_RvdW_cf} can be written as follows~\cite{Fedorov2018}:
\begin{equation}
\label{eq:alpha_RvdW_QDO}
\alpha_1 (\mu\omega,R_{\rm vdW}) = \frac{2^7\cdot (4\pi\varepsilon_0)\, R_{\rm vdW}^7}{(3\hbar/\mu\omega)^2 \, \exp(2\mu\omega R_{\rm vdW}^2/\hbar)} \ .
\end{equation}
The OQDO parametrization imposes that the product $\mu\omega$ in Eq.~\eqref{eq:alpha_RvdW_QDO} delivers the same $R_{\rm vdW}$ as from Eq.~\eqref{eq:dip_pol_RvdW_cf}, for $\alpha_1 (\mu\omega,R_{\rm vdW}) = \alpha_1 (R_{\rm vdW})$.
For simplicity, we rewrite
Eq.~\eqref{eq:alpha_RvdW_QDO} in terms of the dimensionless variable $x$ as
\begin{equation}
\label{eq:TranscEq}
x = a\, e^{b\, x}\ \ ,\ \ \ x = \mu\,\omega\, a_0^2/\hbar = {a_0^2}/{2\sigma^2}\ \ ,
\end{equation}
with the dimensionless coefficients $a$ and $b$ given by
\begin{equation}
\label{eq:Parameters}
a = \frac{3\, \alpha_{\rm fsc}^{\nicefrac 23}}{8\sqrt{2}}\ ,
\ \ b = \frac{R_{\rm vdW}^2}{a_0^2} = \frac{(\alpha_1/4\pi\varepsilon_0)^{\nicefrac 27}}{\alpha_{\rm fsc}^{\nicefrac{8}{21}} 
a_0^{\nicefrac 67}}\ ,
\end{equation}
where we used Eq.~\eqref{eq:dip_pol_RvdW_cf} to express $R_{\rm vdW}$ in terms of $\alpha_1\,$. For all elements in the periodic table, we found that Eq.~\eqref{eq:TranscEq} has two solutions, A and B. This situation is illustrated by the inset of Fig.~\ref{fig:Sigma_Rvdw}b for the case of the argon atom. To rationalize this, it is instructive to consider when Eq.~\eqref{eq:TranscEq} has only one solution, which happens when the polarizability of an atom is equal to the critical
value:
\begin{equation}
\label{eq:limit_case}
{\alpha_1^{\rm (c)}}/{4\pi\varepsilon_0} = \left({8\sqrt{2}}/{3 e}\right)^{_{\nicefrac 72}} \!\! \alpha_{\rm fsc}^{-1}\, a_0^3\, \approx\, 431~ {\rm a.u.}\, ,
\end{equation}
which is greater than the largest known atomic dipole
polarizability ($\alpha_1 \approx 400$~a.u.) of Cs~\cite{Database2018}. It is worth mentioning that the existence of the two solutions extends beyond the employed QDO model. To demonstrate this point, we obtained an analogous result by using the Tang-Toennies potential~\cite{TangToennies2003} with the repulsive interaction treated by the Born-Mayer form (see SM~\cite{SM}).

Since the OQDO frequency is fixed by the second condition of Eqs.~\eqref{eq:fqdo} and \eqref{eq:jqdo}, the solutions A and B for the product $\mu\omega$ differ both in mass and charge, yielding quite different results. First, 
$V_{\mathrm{pol}}^{_{\mathrm{QDO}}}(\mathbf{r})$
constructed from solution B do not resemble DFT potentials, 
while A is in good agreement with them (Fig.~\ref{fig:QDO_response}).
Second, the overlap integral $S=\exp \left( {-\frac{\mu \omega}{2 \hbar} R_{\mathrm{eq}}^2}\right)$ 
between two QDOs at their equilibrium distance $R_{\mathrm{eq}} = 2R_{\mathrm{vdW}}$ is significantly larger
for solution B, which violates the initial assumption used to derive Eq.~\eqref{eq:alpha_RvdW_QDO} that $S$ is small at $R_{\mathrm{eq}}$~\cite{Fedorov2018}.
Third, the QDO length $\sigma$ constructed from solution A follows the same periodic trend as the atomic vdW radii, whereas solution B does not seem to correlate well (see Fig.~\ref{fig:Sigma_Rvdw}a). Therefore, throughout this work we refer to solution A  as the optimized and most promising parametrization. For 102 atoms, the full set of
QDO parameters corresponding to both solutions A and B 
is given in the SM~\cite{SM} together with
the reference values of \{$\alpha_1, C_6$\}. 
Another noteworthy property of the OQDO model (see Fig.~\ref{fig:Sigma_Rvdw}b) is a quasi-linear correlation between the QDO length (model quantity) and the atomic vdW radius (physical observable). In fact, these quantities should be connected \emph{via} the dipole polarizability~\cite{Szabo2022,Fedorov2018}. 
This property is not  captured well by either the FQDO or the JQDO models.

For practical calculations of the vdW energy and constructing predictive force fields, the multipolar contributions associated with the $C_8$ and $C_{10}$ coefficients can become relevant~\cite{XDM,D4,Tao2016}. The available reference data for higher-order molecular dispersion coefficients have significant uncertainties. 
Our careful examination of the literature reporting the reference values of $C_8$ and $C_{10}$ (see \cite{Porsev2003, Porsev2006, JIANG2015, Tao2016} and references therein) identifies uncertainties of up to 20\% for the reference $C_8$ and $C_{10}$ values. Within the QDO formalism, it is straightforward to evaluate these coefficients using closed-form expressions derived by Jones \emph{et al.}~\cite{Jones2013}. In Fig.~\ref{fig:QDO_disp}, we present
the predictions of $C_8$ and $C_{10}$ by FQDO and OQDO models as compared to accurate reference values compiled from the literature~\cite{Porsev2003, Porsev2006, JIANG2015, Tao2016} for a set of 16 atoms (including alkali and alkaline-earth metals and noble gases) and 12 small molecules. We do not include JQDO results in this comparison, since it explicitly uses $C_8$ as an input parameter. Overall, our results show that the OQDO parametrization improves the dispersion coefficients compared to the FQDO one, reducing the MARE from 31\% to 25\% for $C_8$ and from 68\% to 33\% for $C_{10}$ when averaged over all 28 (26 for $C_{10}$) systems. The OQDO model consistently surpasses the FQDO in accuracy for all systems, except for alkaline-earth metals where the FQDO gives more accurate results. Moreover, Fig.~\ref{fig:QDO_disp} shows that the deviations of OQDO dispersion coefficients from the reference values are consistent in terms of their sign and magnitude. Namely, for majority of systems FQDO underestimates $C_8$ and $C_{10}$, but roughly for one third of them the dispersion coefficients are overestimated. The maximal errors of FQDO are observed for Xe both in case of $C_8$ (66\%) and $C_{10}$ (190\%). 
In contrast, OQDO consistently underestimates both dispersion coefficients for all systems, except for $C_8$ of Li as well as $C_{10}$ of Li and Cs, where there is overestimation.
The maximum errors of OQDO are observed for ${\rm CO_2}$ (46\%) in case of $C_8$ and CO (58\%) in case of $C_{10}$, which are significantly smaller than maximum errors of FQDO. 
The consistency of OQDO errors allows a straightforward rescaling of dispersion coefficients: with our best rescaling factors, 1.3 for $C_8$ and 1.5 for $C_{10}$, one can decrease the MARE of OQDO to 15\% and 22\%, respectively, which is not far from the uncertainty of the reference molecular $C_8$ and $C_{10}$ values.
A similar analysis is carried out for multipolar polarizabilities $\alpha_2$ and $\alpha_3$ in the SM~\cite{SM}. In addition, the OQDO model reproduces spatially-distributed polarization potentials rather well for most elements in the periodic table. This highlights the promise of OQDO for constructing \emph{coupled} OQDO models for complex molecules and materials.

Our robust OQDO parametrization for all elements in the periodic table substantially advances the ability to model a wide range of response properties of molecules and materials based on the coupled QDO framework, also paving the way to develop next-generation quantum-mechanical force fields for (bio)molecular simulations.

S.G., A.K., and O.V. contributed equally to this work.
We acknowledge financial support from the European Research Council (ERC Consolidator Grant ``BeStMo''), the Luxembourg National Research Fund \emph{via} FNR CORE Jr project ``PINTA(C17/MS/11686718)'', ``DRIVEN (PRIDE17/12252781)'', and ``ACTIVE (PRIDE19/14063202)''.

\bibliography{biblio}

\begin{thebibliography}{68}
\bibitem [{CCC()}]{CCC}%
  \BibitemOpen
  \href@noop {} {}\bibinfo {note} {The field strength defined this way is
  larger than the one arising in vdW-bonded systems -- the field strengths
  generated by the electron density of free atoms at equilibrium vdW distance
  is on the order of $2\times10^{-6}$~a.u. However, we also assessed a wider
  range of field strengths for three atoms, between $10^{-6}$ and $5 \times
  10^{-2}$ a.u., finding that our conclusions hold for this range as
  well~\cite{SM}.}\BibitemShut {Stop}%
\end{thebibliography}

\end{document}